\documentclass[%
groupedaddress,
preprint,
nofootinbib,
nobibnotes,
 amsmath,amssymb,
 aps,
 prd
]{revtex4}

\usepackage{graphicx}
\usepackage{dcolumn}
\usepackage{bm}

\newcommand{\qed}{\nobreak \ifvmode \relax \else
      \ifdim\lastskip<1.5em \hskip-\lastskip
      \hskip1.5em plus0em minus0.5em \fi \nobreak
      \vrule height0.75em width0.5em depth0.25em\fi}
      
\begin{document}

\preprint{}

\title{ Second Einstein Telescope Mock Science Challenge : \\ Detection of the GW Stochastic Background from Compact Binary Coalescences }
\author{Tania Regimbau}
 \email{regimbau@oca.eu}
\author{Duncan Meacher}
\affiliation{UMR ARTEMIS, CNRS,
University of Nice Sophia-Antipolis, Observatoire de la C\^{o}te d'Azur, CS 34229 F-06304 NICE, France}

\author{Michael Coughlin}
\affiliation{Department of Physics Harvard University, 17 Oxford Street, Cambridge, MA 02138}
\date{\today}

\begin{abstract}
We present the results of the search for an astrophysical gravitational-wave stochastic background during the second Einstein Telescope mock data and science challenge. Assuming that the loudest sources can be detected individually and removed from the data, we show that the residual background can be recovered with an accuracy of $1\%$ with the standard cross-correlation statistic, after correction of a systematic bias due to the non-isotropy of the sources. 
\end{abstract}

\pacs{Valid PACS appear here}
\maketitle

\section{\label{sec:intro} Introduction}
The first generation of gravitational-wave (GW) detectors such as LIGO or Virgo (2002-2013) were able to reach their design sensitivities, demonstrating the feasibility of the experiment. With the second generation, Advanced LIGO \cite{aLIGO} and Advanced Virgo \cite{AdVIRGO}, expected to start collecting data in 2015, we will enter the era of the first GW detections. With a  sensitivity about 10 times better than that of initial LIGO/Virgo, we expect the detection of a few or a few tens of compact binary coalescences (CBC) a year. 

With the third generation european antenna Einstein Telescope (ET) \cite{ET,ETdesign} planned to be operational in $\sim 2025$, GW astronomy will definitely take a step further, with the possibility of detecting a large number of sources (up to $10^4-10^5$ CBC a year) from a large range of processes, such as core collapses to neutron stars or black holes \cite{2005PhRvD..72h4001B,2006PhRvD..73j4024S,2009MNRAS.398..293M,2010MNRAS.409L.132Z}, rotating neutron stars \cite{2001A&A...376..381R,2012PhRvD..86j4007R} including magnetars \cite{2006A&A...447....1R,2011MNRAS.410.2123H,2011MNRAS.411.2549M,2013PhRvD..87d2002W}, phase transition \cite{2009GReGr..41.1389D} or initial instabilities in young neutron stars \cite{1999MNRAS.303..258F,2011ApJ...729...59Z,2004MNRAS.351.1237H,2011ApJ...729...59Z} or compact binary mergers \cite{2011ApJ...739...86Z,2011PhRvD..84h4004R,2011PhRvD..84l4037M,2012PhRvD..85j4024W,2013MNRAS.431..882Z} (see \cite{2011RAA....11..369R} and references therein). 

Besides the emission produced by the coalescence of the nearest binary systems, the superposition of a large number of unresolved sources at high redshifts will produce a background of gravitational waves \cite{2011ApJ...739...86Z,2011PhRvD..84h4004R,2011PhRvD..84l4037M,2012PhRvD..85j4024W,2013MNRAS.431..882Z} that may dominate over the cosmological background in the range $10-1000$ Hz where terrestrial detectors are the most sensitive. The detection of the cosmological background would provide very important constraints on the first instant of the Universe, up to the limits of the Planck era and the Big Bang, while the detection of the astrophysical background would provide crucial information about the star formation history, the mass range of neutron star or black hole progenitors and the rate of compact binary mergers.  

The issue with ET will not be the detection but rather the estimation of the parameters and the interpretation of the results in term of astronomy, cosmology and fundamental physics.
In order to get prepared and test our ability to extract valuable information from the data, we initiated a series of mock data and science challenges, with increasing degree of complexity.

For the first ET mock data and science challenge (ET MDSC1) \cite{2012PhRvD..86l2001R}, we produced one month of simulated data containing simulated gaussian colored noise and the GW signal from a simulated population of double neutron stars in the redshift range $z =0-6$. Using a modified version of the LIGO/Virgo data analysis pipeline IHope \cite{2010PhRvD..82j2001A}, we were able to recover the intrinsic chirp mass and total mass distributions with an error of less than 1\% and 5\% respectively. We also analyzed the data with the standard isotropic cross-correlation (CC) statistic and measured the amplitude of the background  with an accuracy better than 5\%\footnote{Unlike initial LISA which was designed to be a single detector with three arms in a triangle configuration, ET will consist of three nested detectors (six independent arms in total) \cite{ET,ETdesign}, and one can use cross-correlation
methods to extract GW stochastic backgrounds from the instrumental noise.}. Finally, one of our main result was to verify the existence of a null stream  canceling the GW signal and giving a very precise estimate of the noise power spectral density (PSD). By subtracting the null stream from the data, we showed that we could recover the typical shape of the PSD of the GW signal.  

After the success of the first challenge, we extended our data generation package and produced three new sets of data. The first one (ET MDSC2-a) contains all types of stellar compact binary coalescences, composed of two neutron stars (NS-NS), a black hole and a neutron star (BH-NS) or two black holes (BH-BH). The second data set (ET MDSC2-b) contains the population of CBC too faint to be detected individually and which creates a residual GW background. We assume here individual detections can be successfully subtracted  from the data, as it has been done with success for the population of white dwarf binaries in the context of the LISA Mock Data Challenge, using Markov Chain Monte Carlo techniques \cite{2008CQGra..25r4026B}.The third data set (ET MDSC2-c) contains the same population of CBC as ET MDSC2-a, two supernovae and two f-modes,
and also a population of intermediate-mass black hole binary coalescences and intermediate mass ratio inspiral, which could be quite numerous at low frequencies but whose existence has not been confirmed yet \cite{2011GReGr..43..485G}.

In this paper we use the standard cross correlation statistic which is known to be optimal in the case of a Gaussian, isotropic stochastic background  to search for the residual GW background in the second data set ET MDSC2-b. This analysis complement the search for individual CBC (Meacher et al., in preparation), as the majority of the sources contributing to the residual background are at redshift above the detection range. The paper will be organized as follow. In section II we present the CBC population model and summarize briefly the simulation procedure. In section III we discuss the spectral properties of the GW signal in the first and second data sets. In section IV we present the results of the analysis. Section V contains a conclusion and suggestions for further research.

\section{The second ET mock data}   
For each data set, we produced one year of data split into segments of length 2048 s. The data are sampled at 8192 Hz and the minimal frequency was set to 5 Hz (rather than 10 Hz for ET MDSC1) . 
The procedure to generate the simulated data is mostly the same as the one used for ET MDSC1 and is described in detail in \cite{2012PhRvD..86l2001R}.
The main steps are briefly summarized below:

\subsection{Simulation of the Noise}   
\begin{figure}
\includegraphics[angle=0,width=0.9\columnwidth]{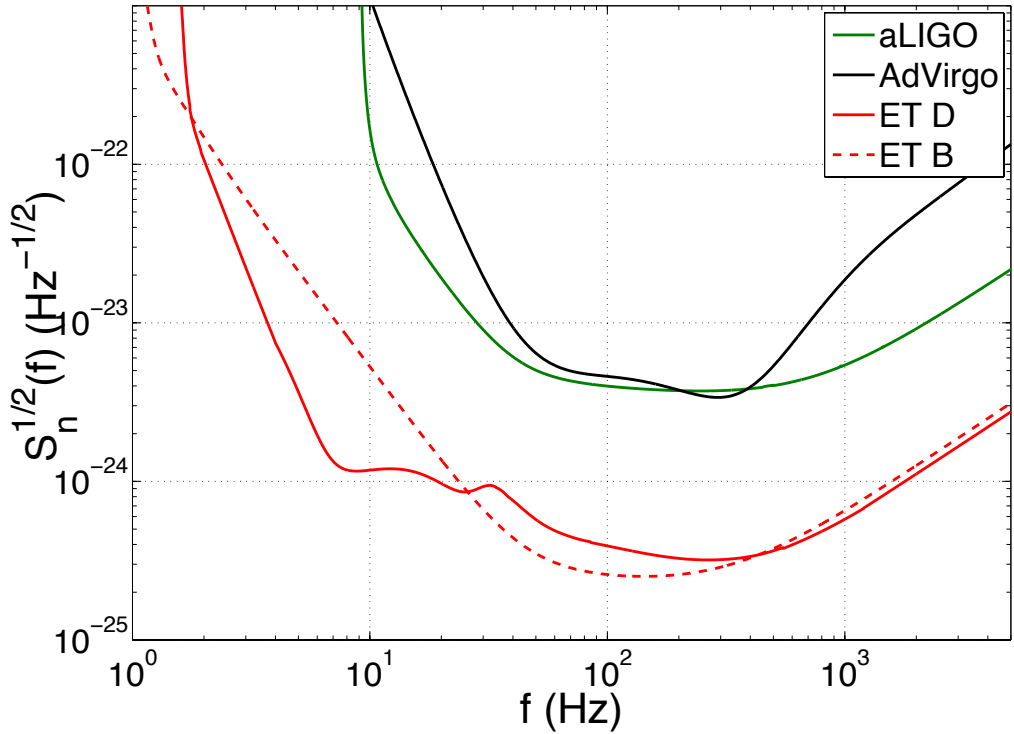}
\caption{The sensitivity curve ET-B (dashed red line) used for the first ET MDSC and and ET-D (continuous red line) used for the second ET MDSC. Advanced LIGO and Virgo noise curves are also shown for comparison.}
\label{fig-noise}
\end{figure}

The Einstein Telescope is envisioned to consist of three independent V-shaped Michelson interferometers with 60 degree opening angles, arranged in a triangle configuration, and placed underground to reduce the influence of seismic noise \cite{ETdesign}.
Assuming there is therefore no instrumental or environmental correlated noise\footnote{Even if the seismic noise will be significantly reduced in ET compared to LIGO or Virgo which are not underground, this assumption may not be realistic as the three ET detectors are nearly co-located. Techniques to identify non-gravitational-wave correlations between a pair of co-located detectors have been developed in the context of the two LIGO detectors at Hanford and could be extended to ET \cite{2008JPhCS.122a2032F}. A more careful study of the effect of environmental noise will be included in future ET mock data and science challenges.}, the noise was simulated independently for each of the three ET detectors E1, E2 and E3, by generating a Gaussian time series with a mean of zero and unit variance. This time series was then transformed into the frequency domain, colored with the noise PSD, and then inverse Fourier transformed. To alleviate the effects of any potential discontinuities across frame files, the noise curve was gradually tapered away to zero below $f_{\min}$, and above $f_{\rm{Nyquist}}/2$. For ET MDSC2, we considered the sensitivity ET-D rather than ET-B for ET MDSC1 (see Fig.~\ref{fig-noise}).

\subsection{Simulation of the GW signal from CBC}

The main improvement compared to the first  ET-MDC is that we generated all types of binaries NS-NS, BH-NS and BH-BH, and did not select the source parameters from simple distributions but used the results of the sophisticated binary evolution code StarTrack \cite{2002ApJ...572..407B,2008ApJS..174..223B,2010ApJ...715L.138B,2012ApJ...759...52D} that provides the masses of the two component $m_1$ and $m_2$ and the delay $t_d$ from formation of the massive binary to final merger. Among their different models, we chose the nominal one \cite{2012ApJ...759...52D} with solar metallicity, small kick velocity and pessimistic common envelop scenario. 

To generate a population of compact binaries, we proceeded as follow for each source :
\begin{itemize}

\item Assuming a Poisson process, the time from the previous coalescence was drawn from an exponential distribution $P(\tau) = \exp(- \tau / \lambda)$. Taking the inverse of the merging rate integrated over all redshifts provided in Table 2 of \cite{2012arXiv1205.4621K}, i.e. 154 929 per year for the nominal model (model BZk), we obtained an average time interval $\lambda=200$ s. 
We then selected the type of the binary using the proportion 84.78\% of NS-NS, 2.09\% of BH-NS and 13.13 \% of BH-BH also provided in Table 2 of \cite{2012arXiv1205.4621K}.
\item the masses $m_1$, $m_2$, and the delay $t_d$ were selected from a list of compact binaries generated by StarTrack. Given the delay and a model for the cosmic star formation rate, we constructed a probability distribution from which the redshift at coalescence $z$ was randomly selected:
\begin{equation}
p(z,t_d) \propto \frac{\dot{\rho}_*(z_f)}{1+z_f} \frac{dV}{dz} dz
\end{equation} 
where $z_f$ is the redshift of formation of the massive binary, $\dot{\rho}_*$ is the star formation rate  and  $\dfrac{dV}{dz}$ is the comoving volume element.The redshifts $z_f$ and $z$ are related by the delay time $t_d$ 
which is the difference in lookback times between $z_f$ and $z$. Following the first ET-MDC, we adopted the star formation rate of \cite{2006ApJ...651..142H} and Lambda-CDM cosmology.
\item the location in the sky $\hat{\Omega}$, the cosinus of the orientation $\iota$, the polarization $\psi$ and the phase at the coalescence $\phi_0$ were drawn from uniform distributions
\item the two polarizations $h_+$ and $h_{\times}$, the antenna pattern functions of the three ET detectors $F^j_+$ and $F^j_{\times}$ ($j=1,2,3$) were calculated, and then the responses $h^j(t)=F^j_+(t) h_+(t)+ F^j_{\times}(t) h_{\times}(t)$ were added to the time series of E1, E2 and E3. In these simulations, we have used so-called TaylorT4 waveforms, 
up to 3.5 post-Newtonian order in phase and the most dominant lowest 
post-Newtonian order term in amplitude for NS-NS and BH-NS, and the EOBNRv2 waveforms including merger and ring down from numerical relativity, up to 4 post-Newtonian order in phase and lowest order in amplitude  for BH-BH \cite{2009PhRvD..80h4043B}.
\end{itemize}

\section{Spectral properties}
The superposition of the GW signal from sources at all redshifts and integrated over all directions of the sky create a background, whose 
spectrum is usually
characterized by the dimensionless energy density parameter \cite{1999PhRvD..59j2001A}:
\begin{equation}
\Omega_{\rm{gw}}(f)=\frac{1}{\rho_c}\frac{d\rho_{\rm{gw}}}{d\ln f},
\end{equation}
where $\rho_{\rm gw}$ is the gravitational energy density and $\rho_c=\dfrac{3c^2H_0^2}{8 \pi G}$
is the critical energy density needed to make the Universe flat today. $G$ is the Newtonian constant, $c$ the speed of light and $H_0$ the Hubble constant.
The GW spectrum from the population of extra-galactic binaries is given by the expression: 
\begin{equation}
\Omega_{\rm{gw}}(f)=\frac{1}{\rho_c c} f F(f).
\label{eq:omega_flux}
\end{equation}
where $F(f)$ is the total flux and $f$ is the observed frequency.
The total flux (in erg Hz$^{-1}$) is the sum of the individual contributions:  
\begin{equation}
F(f)= T^{-1}  \frac{\pi c^3}{2G} f^2 \sum_{k=1}^{N} (\tilde{h}^2_{+,k} + \tilde{h}^2_{\times,k})
\label{eq:flux}
\end{equation}
$N$ is the number of coalescences in the data (the total number of CBC for ET MDSC2-a and the number of undetected sources forming the residual background for ET MDSC2-b). The normalization factor $T^{-1}$ assures that the flux has the correct dimension, $T=1$ yr being the length of the data sample.

In the newtonian regime before the last stable orbit (LSO)  $f_{\rm{lso}} \simeq \dfrac{c^3}{6^{3/2} \pi G M}$, the Fourier transform of $h_+$ and $h_{\times}$ are given by :
\begin{eqnarray}
\tilde{h}_+(f) & = & h_z \, \frac{(1+\cos^2\iota)}{2} \, f^{-7/6},\\
\tilde{h}_\times(f) & = & h_z \, \cos\iota \, f^{-7/6}.
 \label{eq:fourier}
\end{eqnarray}
where
\begin{equation}
h_z = \sqrt\frac{5}{24}\, 
\frac{(G \mathcal{M}(1+z))^{5/6}}{\pi^{2/3} c^{3/2} D_{\rm L}(z)}
\label{eq:hz}
\end{equation}
and where $D_L(z)$ is the luminosity distance at redshift $z$, $\iota$ is the inclination angle, $M=m_1+m_2$ the total mass, $\mathcal{M}=(m_1 m_2)^{3/5}M^{-1/5}$ the chirp mass.
It gives for the energy density parameter :
\begin{equation}
\Omega_{\rm{gw}}(f) = \frac{5 \pi^{2/3} G^{5/3}c^{5/3}}{18 c^3 H_0^2}  f^{2/3} \sum_{k=1}^{N} \frac{(1+z_k)^{5/3} (\mathcal{M}_k)^{5/3}}{D_{\rm L}(z_k)^2}(\frac{(1+\cos^2 \iota_k)^2}{4}+\cos^2\iota_k)
\end{equation}

\begin{figure}
\includegraphics[angle=0,width=0.9\columnwidth]{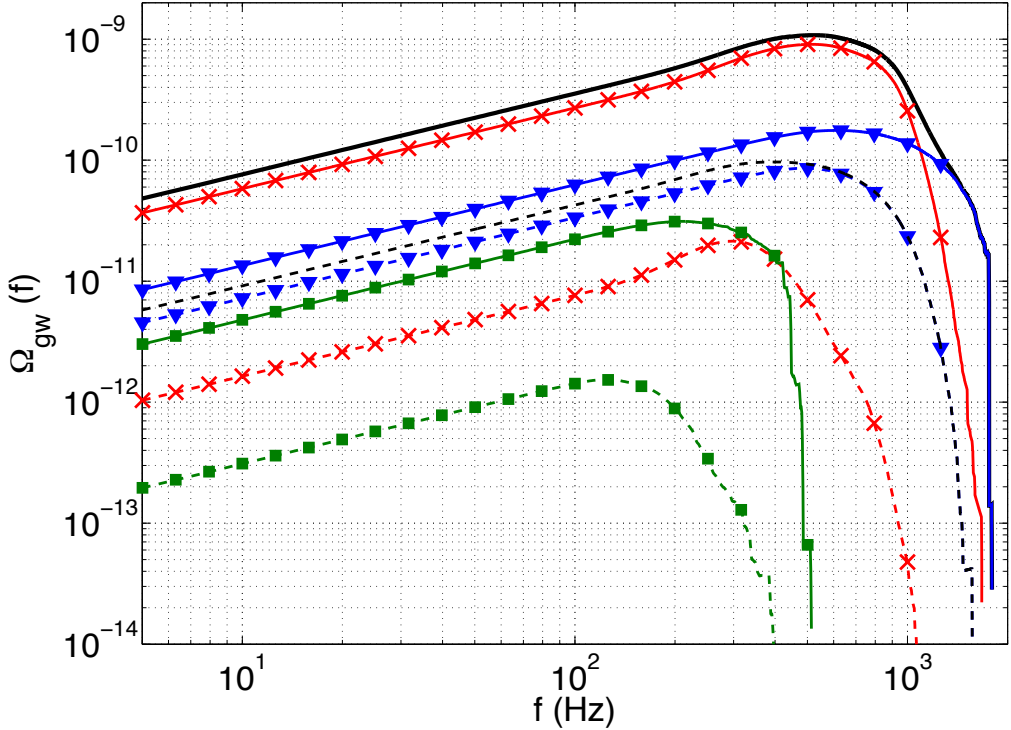}
\caption{$\Omega_{\rm{gw}}$ calculated from the list of CBC sources present in the first (continuous lines) and the second (dashed lines) data sets of ET MDC2, for NS-NS (blue triangles), NS-BH (green squares), BH-BH (red crosses) and the total (black line with no markers}
\label{fig-omega}
\end{figure}

Fig.~\ref{fig-omega} shows $\Omega_{\rm{gw}}$ calculated from the list of CBC sources present in the first (continuous lines) and the second (dashed lines) data sets of ET MDC2, for the different types of binaries and the total.
All the curves present the same characteristic shape : a power law with index $2/3$ at low frequencies corresponding to the inspiral phase, a maximum and a sharp decrease. The BH-BH contribution includes also the merging and ring down phase with a larger power index, that extends after the LSO .

For ET MDC2-a where all the sources are included, the background is dominated by the contribution from BH-BH ($\Omega_{\rm{Ref}} =2.7 \times 10^{-10}$ at $f_{\rm{Ref}}=100$ Hz). The contribution from BH-NS is negligible due to the small rate ($\Omega_{\rm{Ref}} =2.2 \times 10^{-11}$). Even if NS-NS are the most numerous, they are less energetic than BH-BH, and their contribution ($\Omega_{\rm{Ref}} =6.3 \times 10^{-11}$) represents 17\% of the total background ($\Omega_{\rm{Ref}} =3.7 \times 10^{-10}$). 
For ET MDC2-b from which the sources detected individually have been subtracted, the total amplitude drops to $\Omega_{\rm{Ref}} =4.3 \times 10^{-11}$ and it is largely dominated by the contribution from NS-NS ($\Omega_{\rm{Ref}} =3.4 \times 10^{-11}$) which represents 79\% of the total. The contribution from BH-BH, which are the loudest sources, is an order of magnitude below  ($\Omega_{\rm{Ref}} =7.6 \times 10^{-12}$) and two order of magnitude below the BH-BH background in the first data set. The contribution from BH-NS is still negligible with $\Omega_{\rm{Ref}} =1.4 \times 10^{-12}$.

\section{Stochastic Analysis}   
\begin{figure}
\includegraphics[angle=0,width=0.9\columnwidth]{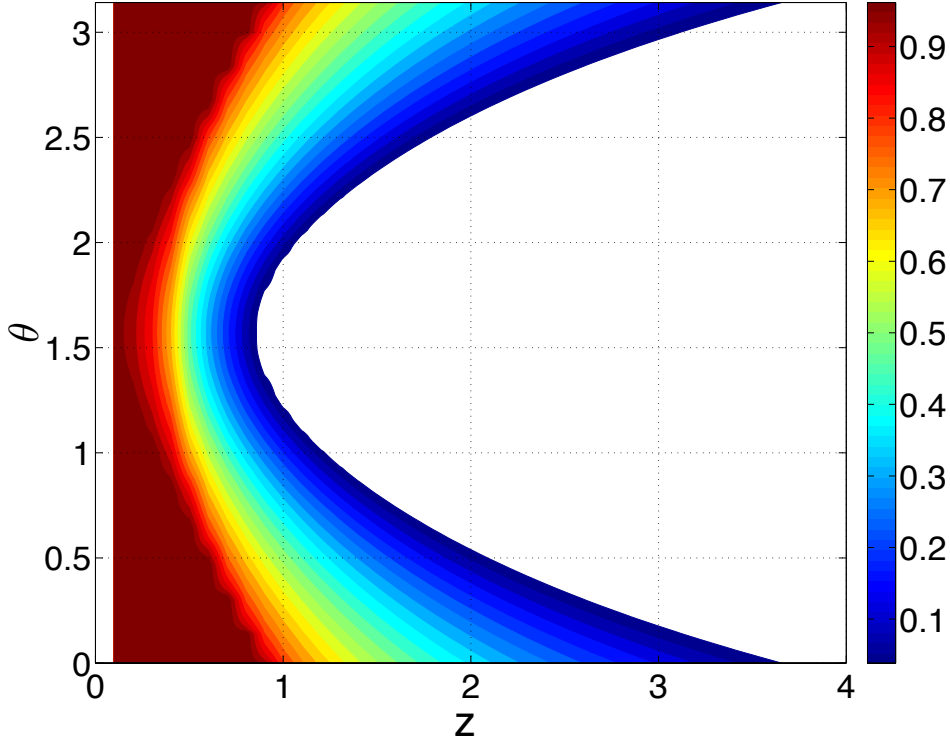}
\caption{Detection efficiency as a function of the polar angle $\theta$ (measured from the zenith direction) for redshifts between $0-10$, for NS-NS sources with masses 1.4+1.4 M$_{\odot}$. The efficiency as a function of the inclination angle $\iota$ has a similar behavior. At a redshift $z=0$ the efficiency is 1 for all the values of $\theta$ and $\iota$ while at the horizon distance of $z \sim 3.8$ only face-on sources located at the zenith or at the nadir are detected. The white area indicates there is no detection.}
\label{fig-effTheta}
\end{figure}

The strategy to search for a Gaussian (or continuous) background, which could be confused 
with the intrinsic noise of a single interferometer, is to cross-correlate measurements of pairs of
detectors. When the background is assumed to be isotropic, unpolarized and stationary, the 
cross-correlation product of detectors $i$ and $j$ is given by 
\cite{1999PhRvD..59j2001A}
\begin{equation}
Y=\int_{0}^\infty \tilde{s}_i^*(f)\tilde{s}_j(f) \tilde{Q}(f)\, {\rm d}f
\label{eq:ccstat}
\end{equation}
and the expected variance, which is dominated by the noise, by
\begin{equation}
\sigma^2_Y \simeq \int_{0}^\infty P_i(f)P_j(f)|\tilde{Q}(f)|^2\,{\rm d}f,
\label{eq:ccvar}
\end{equation}
where
\begin{equation}
\tilde{Q}(f)\propto \frac{\gamma_{ij} (f) \Omega_{\rm gw}(f)}{f^3 P_i(f) P_j(f)}
\end{equation}
is a filter that maximizes the signal-to-noise ratio, 
\begin{equation}
\mathrm{SNR} =\frac{3 H_0^2}{4 \pi^2} \sqrt{2T} \left[ \int_0^\infty
df \frac{\gamma_{ij}^2(f)\Omega_{\rm gw}^2(f)}{f^6 P_i(f)P_j(f)} \right]^{1/2}
\label{eq:snrCC}
\end{equation}
In the above equations, $P_i$ 
and $P_j$ are the one-sided power spectral noise densities of the two detectors and $\gamma_{ij}$ is the normalized isotropic overlap reduction function (ORF), characterizing the loss of 
sensitivity due to the separation and the relative orientation of the detectors for sources isotropically distributed in the sky \cite{1993PhRvD..48.2389F,1992PhRvD..46.5250C}. 
\begin{equation}
\gamma_{ij}(f) = \frac{5 \sin^2(\gamma)}{8\pi} \int d\hat{k} \, e^{i 2\pi f \hat
k \cdot \Delta \vec{x}_{ij}/c
 } ( F^i_+  F^j_+  + F^i_\times  F^j_\times) ,
\label{eq:gamma-iso}
\end{equation}
where $\Delta \vec{x}_{ij}$ is the separation vector between the
vertices of the two detectors and the product $\hat
k \cdot \Delta \vec{x}_{ij}/c$ in the exponential, the time delay for a wave arriving from direction $\hat{k}$. The normalization ensures that $\gamma_{ij} = 1$ for co-located and co-aligned L-shaped detectors. For two V-shaped ($\alpha = \pi/3$) ET detectors separated 
by $\beta = 2 \pi/3$ degrees and with $f \Delta t <<1 $ , $\gamma_{ij} = \sin^2(\alpha) \cos(2 \beta) = -3/8=-0.375$. 
In a recent paper Meacher et al. \cite{duncan} derived a general expression of the ORF, valid for any distribution in the parameter space of sources contributing to the background. For ET it writes :
\begin{equation}
\hat{\gamma_{ij}}(f) = \frac{3}{10} \frac{ \sum\limits_{n=1}^N h_{z, k}^2 F_{ij,k}}{N \left\langle h_z^2 \left( \dfrac{\left(1+\cos^2\iota \right)^2}{4} + \cos^2\iota \right) \right\rangle}
 \label{eq:gamma}
\end{equation}
with
\begin{equation}
h_{z,k} = \sqrt\frac{5}{24}\, 
\frac{(G \mathcal{M}_k(1+z_k))^{5/6}}{\pi^{2/3} c^{3/2} D_{\rm L}(z_k)}
\end{equation}
and
\begin{eqnarray}
F_{ij,k} &=& \left[ \frac{\left(1+\cos^2\iota_k \right)^2}{4}  F^i_{+,k} F^j_{+,k} + \cos^2\iota_k F^i_{+,k}\times F^j_{\times,k} + \right. 
 \left.  \frac{\left(1+\cos^2\iota_k \right)\cos\iota_k}{2}  \left(F^i_{+,k} F^j_{\times,k} + F^i_{\times,k} F^j_{+,k} \right) \right]  \notag \\
 &\simeq& \left[ \frac{\left(1+\cos^2\iota_k \right)^2}{4}  F^i_{+,k} F^j_{+,k} + \cos^2\iota_k F^i_{\times,k} F^j_{\times,k}  \right]
\end{eqnarray}
The angle brackets $<>$ in Eq.~\ref{eq:gamma} indicate an average over all the sources present in the data, which means in the case of the residual background, all the CBC sources with a signal-to-noise ratio smaller that 8. For ET MDC2-b we obtain $\hat{\gamma}_{12}(f)=-0.283$, $\hat{\gamma}_{13}(f)=-0.284$ and $\hat{\gamma}_{23}(f)=-0.281$, so a factor of about 1.3  smaller than the isotropic value. When the detected sources are removed, there is a selection effect that affects the distribution of the parameters (in particular the isotropy and the uniform orientation). After the detector horizon (the maximal distance at which a source can be detected) all the sources contribute to the background, but at close redshifts, only poorly oriented or located sources contribute (see Fig.~\ref{fig-effTheta}).  
This effect is not negligible and has to be corrected in order to avoid a systematic bias in the analysis. A priori the distribution of the sources in the parameter space is not known and neither the correction factor. However, for a narrow distribution of the masses (which is the case here since the background is largely dominated by the BNS population), it only depends on the distribution in redshift and the average chirp mass and one can easily obtain the expected value (in the limit $N>>1$) from the star formation rate. Doing this, we obtained a correction of $\sim 1.25$, in agreement with our results for ET MDC2-b with a precision better than $4\%$.

\begin{figure}
\centering
\includegraphics[angle=000,width=0.9\textwidth]{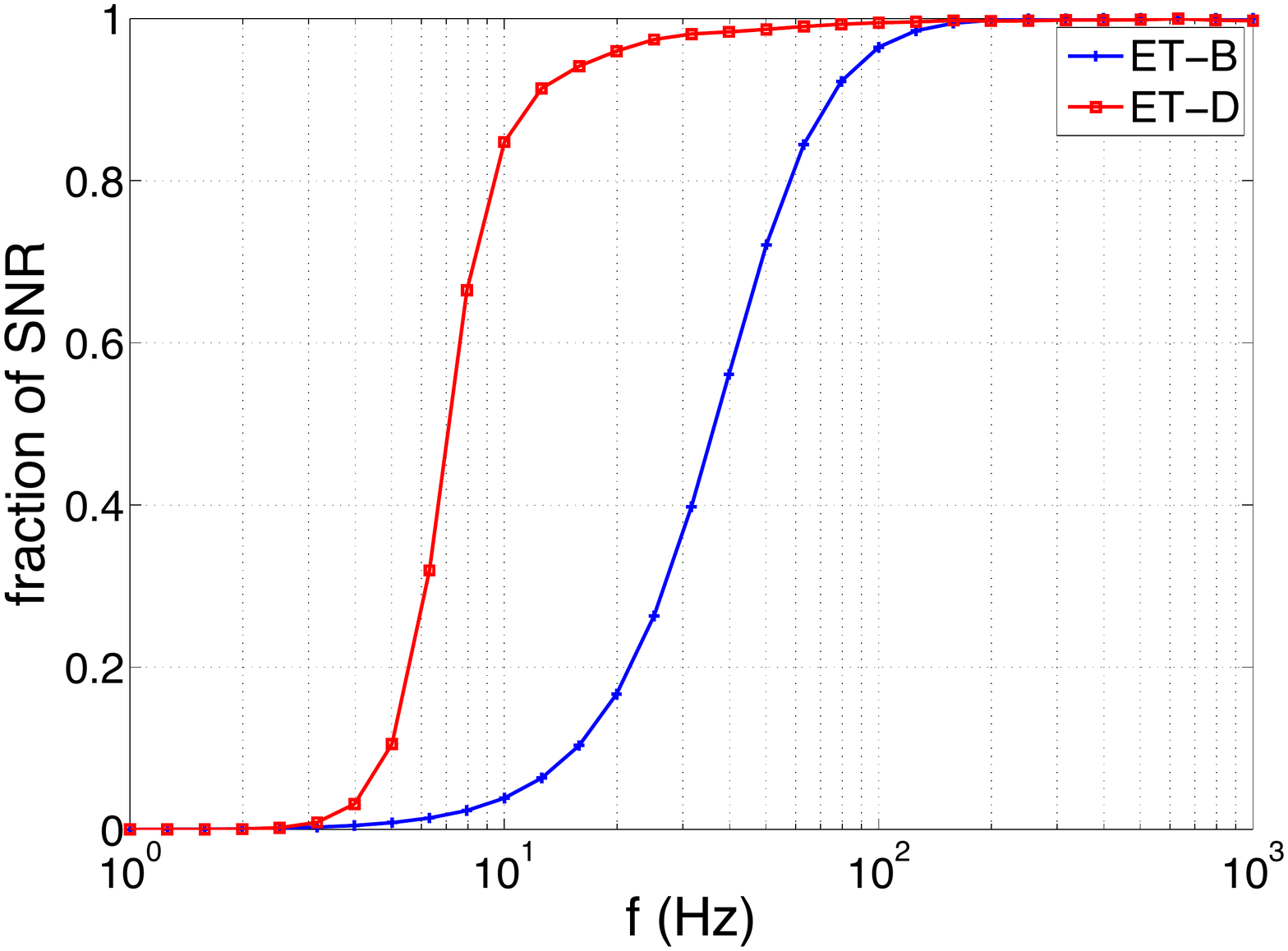}
\caption{Contribution to the SNR of frequencies $<f$, for ET-D (ET MDC2) and ET-B (ET MDC1) \cite{2012arXiv1205.4621K}}
\label{fig-snr_fraction}
\end{figure}

We analyzed the data with the cross-correlation code developed by the LIGO stochastic 
group. The data were split into $N=529067$ segments of length $T_{\mathrm{seg}}=60$\,s, and for 
each segment the cross-correlation product and the theoretical variance were calculated 
using a template $\Omega_{\mathrm{gw}} \sim f^{2/3}$ in the range $5-150$\,Hz, where we have more than $99\%$ of the SNR (see Fig.~\ref{fig-snr_fraction}) \cite{2012arXiv1205.4621K,2013MNRAS.431..882Z}.
The frequency 
resolution of our analysis was 0.25\,Hz. The final point estimate at 10\,Hz is given by 
\cite{1999PhRvD..59j2001A}
\begin{equation}
\hat{\Omega}_{\mathrm{gw}} = \frac{\sum_i Y_i\, \sigma_{Y,i}^{-2}}{T_{\rm{seg}} \sum_i \sigma_{Y,i}^{-2}} 
\end{equation}
where $Y_i$ and $\sigma_{Y,i}^2$ are the cross-correlations and variances calculated for
each segment using Eq.~\eqref{eq:ccstat}, \eqref{eq:ccvar} respectively.
The standard error on this estimate is given by
\begin{equation}
\sigma_{\Omega_{\mathrm{gw}} }= T_{\rm{seg}}^{-1} \left[\sum_i \sigma_{Y,i}^{-2}\right]^{-1/2}  .
\end{equation}

The isotropic analysis gives a point estimate of  $3.21 \times 10^{-12}$ for E1-E2, $3.18 \times 10^{-12}$ for E1-E3, $3.22 \times 10^{-12}$ for E2-E3, so an average of $\sim 3.2 \times 10^{-12}$, with error $\sigma_{\Omega_{\mathrm{gw}} } =  4.4 \times 10^{-12}$ for the three pairs. Applying the correction factor of $1.3$ derived above for the ORF, we found a point estimate at 100\,Hz of $ 4.26\times 10^{-12}$ for the pair E1-E2, $4.20 \times 10^{-12}$ for E1-E3 
and $4.3 \times 10^{-12}$ for E2-E3 . The average is $\hat{\Omega}_{\mathrm{gw}} \sim 4.25 \times 10^{-12}$, which corresponds to the analytical 
expectation of $\sim 4.3 \times 10^{-12}$ with a precision of about $1 \%$.

\section{\label{sec:conclusion} Conclusion}
In this paper we reported on the analysis of the second Einstein Telescope mock data and science challenge, searching for the residual GW background resulting from the superposition of all the CBC sources that are too faint to be detected individually. We used the standard cross correlation statistic which is known to be optimal in the case of a Gaussian, isotropic stochastic background. Confirming the results of the ET MDSC1 and the recent work of Meacher et al., we obtained that  the non continuity or non gaussianity of the background \cite{2011PhRvD..84h4004R,2012PhRvD..85j4024W,2012arXiv1205.4621K,2012PhRvD..86l2001R} does not significantly affect the analysis (what's important is the total number of sources and not whether they overlap or not). But because of the GW selection effect that favored the detection of the best oriented and located sources, especially at larger redshift, the assumption of an isotropic stochastic background is not verified and the estimate given by the standard  cross correlation statistic presents a systematic bias in the case of the residual background. Deriving a correction for the overlap reduction function we obtained a point estimate that agrees with the expected value with a precision $<1\%$. 
The detection of the residual background would have very important consequences in cosmology and astrophysics as it would probe the high redshift population, complementing individual detections at smaller redshift. The residual background from CBC may dominate in the frequency band of ET. In future ET MDSC, we will investigate how one can use the non Gaussian signature to separate this background or foreground and recover the cosmological background.
 
\begin{acknowledgments}
M. Coughlin is supported by National Science
Foundation Graduate Research Fellowship Program, under
NSF grant number DGE 1144152.
\end{acknowledgments}

\bibliography{biblio}

\begin{thebibliography}{40}
\expandafter\ifx\csname natexlab\endcsname\relax\def\natexlab#1{#1}\fi
\expandafter\ifx\csname bibnamefont\endcsname\relax
  \def\bibnamefont#1{#1}\fi
\expandafter\ifx\csname bibfnamefont\endcsname\relax
  \def\bibfnamefont#1{#1}\fi
\expandafter\ifx\csname citenamefont\endcsname\relax
  \def\citenamefont#1{#1}\fi
\expandafter\ifx\csname url\endcsname\relax
  \def\url#1{\texttt{#1}}\fi
\expandafter\ifx\csname urlprefix\endcsname\relax\def\urlprefix{URL }\fi
\providecommand{\bibinfo}[2]{#2}
\providecommand{\eprint}[2][]{\url{#2}}

\bibitem[{\citenamefont{the Advanced LIGO~Team}(2007)}]{aLIGO}
\bibinfo{author}{\bibnamefont{the Advanced LIGO~Team}} (\bibinfo{year}{2007}),
  \urlprefix\url{https://dcc.ligo.org/cgi-bin/DocDB/ShowDocument?docid=m060056}.

\bibitem[{\citenamefont{Losurdo and the Advanced Virgo~Team}(2007)}]{AdVIRGO}
\bibinfo{author}{\bibfnamefont{G.}~\bibnamefont{Losurdo}} \bibnamefont{and}
  \bibinfo{author}{\bibnamefont{the Advanced Virgo~Team}}
  (\bibinfo{year}{2007}), \urlprefix\url{https://tds.ego-gw.it/ql/?c=1900}.

\bibitem[{\citenamefont{Punturo et~al.}(2010)}]{ET}
\bibinfo{author}{\bibfnamefont{M.}~\bibnamefont{Punturo}} \bibnamefont{et~al.},
  \bibinfo{journal}{Classical and Quantum Gravity}
  \textbf{\bibinfo{volume}{27}}, \bibinfo{pages}{194002}
  (\bibinfo{year}{2010}),
  \urlprefix\url{http://stacks.iop.org/0264-9381/27/i=19/a=194002}.

\bibitem[{\citenamefont{{Aberthany} and et~al.}(2012)}]{ETdesign}
\bibinfo{author}{\bibfnamefont{M.}~\bibnamefont{{Aberthany}}} \bibnamefont{and}
  \bibinfo{author}{\bibnamefont{et~al.}}, \bibinfo{journal}{European
  Gravitational Observatory document number ET-0106A-10,
  http://www.et-gw.eu/etdsdocument}  (\bibinfo{year}{2012}).

\bibitem[{\citenamefont{{Buonanno} et~al.}(2005)\citenamefont{{Buonanno},
  {Sigl}, {Raffelt}, {Janka}, and {M{\"u}ller}}}]{2005PhRvD..72h4001B}
\bibinfo{author}{\bibfnamefont{A.}~\bibnamefont{{Buonanno}}},
  \bibinfo{author}{\bibfnamefont{G.}~\bibnamefont{{Sigl}}},
  \bibinfo{author}{\bibfnamefont{G.~G.} \bibnamefont{{Raffelt}}},
  \bibinfo{author}{\bibfnamefont{H.-T.} \bibnamefont{{Janka}}},
  \bibnamefont{and}
  \bibinfo{author}{\bibfnamefont{E.}~\bibnamefont{{M{\"u}ller}}},
  \bibinfo{journal}{\prd} \textbf{\bibinfo{volume}{72}}, \bibinfo{eid}{084001}
  (\bibinfo{year}{2005}), \eprint{astro-ph/0412277}.

\bibitem[{\citenamefont{{Sandick} et~al.}(2006)\citenamefont{{Sandick},
  {Olive}, {Daigne}, and {Vangioni}}}]{2006PhRvD..73j4024S}
\bibinfo{author}{\bibfnamefont{P.}~\bibnamefont{{Sandick}}},
  \bibinfo{author}{\bibfnamefont{K.~A.} \bibnamefont{{Olive}}},
  \bibinfo{author}{\bibfnamefont{F.}~\bibnamefont{{Daigne}}}, \bibnamefont{and}
  \bibinfo{author}{\bibfnamefont{E.}~\bibnamefont{{Vangioni}}},
  \bibinfo{journal}{\prd} \textbf{\bibinfo{volume}{73}}, \bibinfo{eid}{104024}
  (\bibinfo{year}{2006}), \eprint{astro-ph/0603544}.

\bibitem[{\citenamefont{{Marassi} et~al.}(2009)\citenamefont{{Marassi},
  {Schneider}, and {Ferrari}}}]{2009MNRAS.398..293M}
\bibinfo{author}{\bibfnamefont{S.}~\bibnamefont{{Marassi}}},
  \bibinfo{author}{\bibfnamefont{R.}~\bibnamefont{{Schneider}}},
  \bibnamefont{and}
  \bibinfo{author}{\bibfnamefont{V.}~\bibnamefont{{Ferrari}}},
  \bibinfo{journal}{Monthly Notices of the Royal astronomical Society}
  \textbf{\bibinfo{volume}{398}}, \bibinfo{pages}{293} (\bibinfo{year}{2009}),
  \eprint{0906.0461}.

\bibitem[{\citenamefont{{Zhu} et~al.}(2010)\citenamefont{{Zhu}, {Howell}, and
  {Blair}}}]{2010MNRAS.409L.132Z}
\bibinfo{author}{\bibfnamefont{X.-J.} \bibnamefont{{Zhu}}},
  \bibinfo{author}{\bibfnamefont{E.}~\bibnamefont{{Howell}}}, \bibnamefont{and}
  \bibinfo{author}{\bibfnamefont{D.}~\bibnamefont{{Blair}}},
  \bibinfo{journal}{Monthly Notices of the Royal astronomical Society}
  \textbf{\bibinfo{volume}{409}}, \bibinfo{pages}{L132} (\bibinfo{year}{2010}),
  \eprint{1008.0472}.

\bibitem[{\citenamefont{{Regimbau} and {de Freitas
  Pacheco}}(2001)}]{2001A&A...376..381R}
\bibinfo{author}{\bibfnamefont{T.}~\bibnamefont{{Regimbau}}} \bibnamefont{and}
  \bibinfo{author}{\bibfnamefont{J.~A.} \bibnamefont{{de Freitas Pacheco}}},
  \bibinfo{journal}{Astronomy and Astrophysics} \textbf{\bibinfo{volume}{376}},
  \bibinfo{pages}{381} (\bibinfo{year}{2001}), \eprint{astro-ph/0105260}.

\bibitem[{\citenamefont{{Rosado}}(2012)}]{2012PhRvD..86j4007R}
\bibinfo{author}{\bibfnamefont{P.~A.} \bibnamefont{{Rosado}}},
  \bibinfo{journal}{\prd} \textbf{\bibinfo{volume}{86}}, \bibinfo{eid}{104007}
  (\bibinfo{year}{2012}), \eprint{1206.1330}.

\bibitem[{\citenamefont{{Regimbau} and {de Freitas
  Pacheco}}(2006)}]{2006A&A...447....1R}
\bibinfo{author}{\bibfnamefont{T.}~\bibnamefont{{Regimbau}}} \bibnamefont{and}
  \bibinfo{author}{\bibfnamefont{J.~A.} \bibnamefont{{de Freitas Pacheco}}},
  \bibinfo{journal}{Astronomy and Astrophysics} \textbf{\bibinfo{volume}{447}},
  \bibinfo{pages}{1} (\bibinfo{year}{2006}), \eprint{astro-ph/0509880}.

\bibitem[{\citenamefont{{Howell} et~al.}(2011)\citenamefont{{Howell},
  {Regimbau}, {Corsi}, {Coward}, and {Burman}}}]{2011MNRAS.410.2123H}
\bibinfo{author}{\bibfnamefont{E.}~\bibnamefont{{Howell}}},
  \bibinfo{author}{\bibfnamefont{T.}~\bibnamefont{{Regimbau}}},
  \bibinfo{author}{\bibfnamefont{A.}~\bibnamefont{{Corsi}}},
  \bibinfo{author}{\bibfnamefont{D.}~\bibnamefont{{Coward}}}, \bibnamefont{and}
  \bibinfo{author}{\bibfnamefont{R.}~\bibnamefont{{Burman}}},
  \bibinfo{journal}{Monthly Notices of the Royal astronomical Society}
  \textbf{\bibinfo{volume}{410}}, \bibinfo{pages}{2123} (\bibinfo{year}{2011}),
  \eprint{1008.3941}.

\bibitem[{\citenamefont{{Marassi}
  et~al.}(2011{\natexlab{a}})\citenamefont{{Marassi}, {Ciolfi}, {Schneider},
  {Stella}, and {Ferrari}}}]{2011MNRAS.411.2549M}
\bibinfo{author}{\bibfnamefont{S.}~\bibnamefont{{Marassi}}},
  \bibinfo{author}{\bibfnamefont{R.}~\bibnamefont{{Ciolfi}}},
  \bibinfo{author}{\bibfnamefont{R.}~\bibnamefont{{Schneider}}},
  \bibinfo{author}{\bibfnamefont{L.}~\bibnamefont{{Stella}}}, \bibnamefont{and}
  \bibinfo{author}{\bibfnamefont{V.}~\bibnamefont{{Ferrari}}},
  \bibinfo{journal}{Monthly Notices of the Royal astronomical Society}
  \textbf{\bibinfo{volume}{411}}, \bibinfo{pages}{2549}
  (\bibinfo{year}{2011}{\natexlab{a}}), \eprint{1009.1240}.

\bibitem[{\citenamefont{{Wu} et~al.}(2013)\citenamefont{{Wu}, {Mandic}, and
  {Regimbau}}}]{2013PhRvD..87d2002W}
\bibinfo{author}{\bibfnamefont{C.-J.} \bibnamefont{{Wu}}},
  \bibinfo{author}{\bibfnamefont{V.}~\bibnamefont{{Mandic}}}, \bibnamefont{and}
  \bibinfo{author}{\bibfnamefont{T.}~\bibnamefont{{Regimbau}}},
  \bibinfo{journal}{\prd} \textbf{\bibinfo{volume}{87}}, \bibinfo{eid}{042002}
  (\bibinfo{year}{2013}).

\bibitem[{\citenamefont{{de Araujo} and
  {Marranghello}}(2009)}]{2009GReGr..41.1389D}
\bibinfo{author}{\bibfnamefont{J.~C.~N.} \bibnamefont{{de Araujo}}}
  \bibnamefont{and} \bibinfo{author}{\bibfnamefont{G.~F.}
  \bibnamefont{{Marranghello}}}, \bibinfo{journal}{General Relativity and
  Gravitation} \textbf{\bibinfo{volume}{41}}, \bibinfo{pages}{1389}
  (\bibinfo{year}{2009}), \eprint{0910.0424}.

\bibitem[{\citenamefont{{Ferrari} et~al.}(1999)\citenamefont{{Ferrari},
  {Matarrese}, and {Schneider}}}]{1999MNRAS.303..258F}
\bibinfo{author}{\bibfnamefont{V.}~\bibnamefont{{Ferrari}}},
  \bibinfo{author}{\bibfnamefont{S.}~\bibnamefont{{Matarrese}}},
  \bibnamefont{and}
  \bibinfo{author}{\bibfnamefont{R.}~\bibnamefont{{Schneider}}},
  \bibinfo{journal}{Monthly Notices of the Royal astronomical Society}
  \textbf{\bibinfo{volume}{303}}, \bibinfo{pages}{258} (\bibinfo{year}{1999}),
  \eprint{astro-ph/9806357}.

\bibitem[{\citenamefont{{Zhu} et~al.}(2011{\natexlab{a}})\citenamefont{{Zhu},
  {Fan}, and {Zhu}}}]{2011ApJ...729...59Z}
\bibinfo{author}{\bibfnamefont{X.-J.} \bibnamefont{{Zhu}}},
  \bibinfo{author}{\bibfnamefont{X.-L.} \bibnamefont{{Fan}}}, \bibnamefont{and}
  \bibinfo{author}{\bibfnamefont{Z.-H.} \bibnamefont{{Zhu}}},
  \bibinfo{journal}{\apj} \textbf{\bibinfo{volume}{729}}, \bibinfo{eid}{59}
  (\bibinfo{year}{2011}{\natexlab{a}}), \eprint{1102.2786}.

\bibitem[{\citenamefont{{Howell} et~al.}(2004)\citenamefont{{Howell}, {Coward},
  {Burman}, {Blair}, and {Gilmore}}}]{2004MNRAS.351.1237H}
\bibinfo{author}{\bibfnamefont{E.}~\bibnamefont{{Howell}}},
  \bibinfo{author}{\bibfnamefont{D.}~\bibnamefont{{Coward}}},
  \bibinfo{author}{\bibfnamefont{R.}~\bibnamefont{{Burman}}},
  \bibinfo{author}{\bibfnamefont{D.}~\bibnamefont{{Blair}}}, \bibnamefont{and}
  \bibinfo{author}{\bibfnamefont{J.}~\bibnamefont{{Gilmore}}},
  \bibinfo{journal}{Monthly Notices of the Royal astronomical Society}
  \textbf{\bibinfo{volume}{351}}, \bibinfo{pages}{1237} (\bibinfo{year}{2004}).

\bibitem[{\citenamefont{{Zhu} et~al.}(2011{\natexlab{b}})\citenamefont{{Zhu},
  {Howell}, {Regimbau}, {Blair}, and {Zhu}}}]{2011ApJ...739...86Z}
\bibinfo{author}{\bibfnamefont{X.-J.} \bibnamefont{{Zhu}}},
  \bibinfo{author}{\bibfnamefont{E.}~\bibnamefont{{Howell}}},
  \bibinfo{author}{\bibfnamefont{T.}~\bibnamefont{{Regimbau}}},
  \bibinfo{author}{\bibfnamefont{D.}~\bibnamefont{{Blair}}}, \bibnamefont{and}
  \bibinfo{author}{\bibfnamefont{Z.-H.} \bibnamefont{{Zhu}}},
  \bibinfo{journal}{\apj} \textbf{\bibinfo{volume}{739}}, \bibinfo{eid}{86}
  (\bibinfo{year}{2011}{\natexlab{b}}), \eprint{1104.3565}.

\bibitem[{\citenamefont{{Rosado}}(2011)}]{2011PhRvD..84h4004R}
\bibinfo{author}{\bibfnamefont{P.~A.} \bibnamefont{{Rosado}}},
  \bibinfo{journal}{\prd} \textbf{\bibinfo{volume}{84}}, \bibinfo{eid}{084004}
  (\bibinfo{year}{2011}), \eprint{1106.5795}.

\bibitem[{\citenamefont{{Marassi}
  et~al.}(2011{\natexlab{b}})\citenamefont{{Marassi}, {Schneider}, {Corvino},
  {Ferrari}, and {Zwart}}}]{2011PhRvD..84l4037M}
\bibinfo{author}{\bibfnamefont{S.}~\bibnamefont{{Marassi}}},
  \bibinfo{author}{\bibfnamefont{R.}~\bibnamefont{{Schneider}}},
  \bibinfo{author}{\bibfnamefont{G.}~\bibnamefont{{Corvino}}},
  \bibinfo{author}{\bibfnamefont{V.}~\bibnamefont{{Ferrari}}},
  \bibnamefont{and} \bibinfo{author}{\bibfnamefont{S.~P.}
  \bibnamefont{{Zwart}}}, \bibinfo{journal}{\prd}
  \textbf{\bibinfo{volume}{84}}, \bibinfo{eid}{124037}
  (\bibinfo{year}{2011}{\natexlab{b}}), \eprint{1111.6125}.

\bibitem[{\citenamefont{{Wu} et~al.}(2012)\citenamefont{{Wu}, {Mandic}, and
  {Regimbau}}}]{2012PhRvD..85j4024W}
\bibinfo{author}{\bibfnamefont{C.}~\bibnamefont{{Wu}}},
  \bibinfo{author}{\bibfnamefont{V.}~\bibnamefont{{Mandic}}}, \bibnamefont{and}
  \bibinfo{author}{\bibfnamefont{T.}~\bibnamefont{{Regimbau}}},
  \bibinfo{journal}{\prd} \textbf{\bibinfo{volume}{85}}, \bibinfo{eid}{104024}
  (\bibinfo{year}{2012}), \eprint{1112.1898}.

\bibitem[{\citenamefont{{Zhu} et~al.}(2013)\citenamefont{{Zhu}, {Howell},
  {Blair}, and {Zhu}}}]{2013MNRAS.431..882Z}
\bibinfo{author}{\bibfnamefont{X.-J.} \bibnamefont{{Zhu}}},
  \bibinfo{author}{\bibfnamefont{E.~J.} \bibnamefont{{Howell}}},
  \bibinfo{author}{\bibfnamefont{D.~G.} \bibnamefont{{Blair}}},
  \bibnamefont{and} \bibinfo{author}{\bibfnamefont{Z.-H.} \bibnamefont{{Zhu}}},
  \bibinfo{journal}{Monthly Notices of the Royal astronomical Society}
  \textbf{\bibinfo{volume}{431}}, \bibinfo{pages}{882} (\bibinfo{year}{2013}),
  \eprint{1209.0595}.

\bibitem[{\citenamefont{{Regimbau}}(2011)}]{2011RAA....11..369R}
\bibinfo{author}{\bibfnamefont{T.}~\bibnamefont{{Regimbau}}},
  \bibinfo{journal}{Research in Astronomy and Astrophysics}
  \textbf{\bibinfo{volume}{11}}, \bibinfo{pages}{369} (\bibinfo{year}{2011}),
  \eprint{1101.2762}.

\bibitem[{\citenamefont{{Regimbau} et~al.}(2012)\citenamefont{{Regimbau},
  {Dent}, {Del Pozzo}, {Giampanis}, {Li}, {Robinson}, {Van Den Broeck},
  {Meacher}, {Rodriguez}, {Sathyaprakash} et~al.}}]{2012PhRvD..86l2001R}
\bibinfo{author}{\bibfnamefont{T.}~\bibnamefont{{Regimbau}}},
  \bibinfo{author}{\bibfnamefont{T.}~\bibnamefont{{Dent}}},
  \bibinfo{author}{\bibfnamefont{W.}~\bibnamefont{{Del Pozzo}}},
  \bibinfo{author}{\bibfnamefont{S.}~\bibnamefont{{Giampanis}}},
  \bibinfo{author}{\bibfnamefont{T.~G.~F.} \bibnamefont{{Li}}},
  \bibinfo{author}{\bibfnamefont{C.}~\bibnamefont{{Robinson}}},
  \bibinfo{author}{\bibfnamefont{C.}~\bibnamefont{{Van Den Broeck}}},
  \bibinfo{author}{\bibfnamefont{D.}~\bibnamefont{{Meacher}}},
  \bibinfo{author}{\bibfnamefont{C.}~\bibnamefont{{Rodriguez}}},
  \bibinfo{author}{\bibfnamefont{B.~S.} \bibnamefont{{Sathyaprakash}}},
  \bibnamefont{et~al.}, \bibinfo{journal}{\prd} \textbf{\bibinfo{volume}{86}},
  \bibinfo{eid}{122001} (\bibinfo{year}{2012}), \eprint{1201.3563}.

\bibitem[{\citenamefont{{Abadie} et~al.}(2010)\citenamefont{{Abadie}, {Abbott},
  {Abbott}, {Abernathy}, {Accadia}, {Acernese}, {Adams}, {Adhikari}, {Ajith},
  {Allen} et~al.}}]{2010PhRvD..82j2001A}
\bibinfo{author}{\bibfnamefont{J.}~\bibnamefont{{Abadie}}},
  \bibinfo{author}{\bibfnamefont{B.~P.} \bibnamefont{{Abbott}}},
  \bibinfo{author}{\bibfnamefont{R.}~\bibnamefont{{Abbott}}},
  \bibinfo{author}{\bibfnamefont{M.}~\bibnamefont{{Abernathy}}},
  \bibinfo{author}{\bibfnamefont{T.}~\bibnamefont{{Accadia}}},
  \bibinfo{author}{\bibfnamefont{F.}~\bibnamefont{{Acernese}}},
  \bibinfo{author}{\bibfnamefont{C.}~\bibnamefont{{Adams}}},
  \bibinfo{author}{\bibfnamefont{R.}~\bibnamefont{{Adhikari}}},
  \bibinfo{author}{\bibfnamefont{P.}~\bibnamefont{{Ajith}}},
  \bibinfo{author}{\bibfnamefont{B.}~\bibnamefont{{Allen}}},
  \bibnamefont{et~al.}, \bibinfo{journal}{\prd} \textbf{\bibinfo{volume}{82}},
  \bibinfo{eid}{102001} (\bibinfo{year}{2010}), \eprint{1005.4655}.

\bibitem[{\citenamefont{{Babak} et~al.}(2008)\citenamefont{{Babak}, {Baker},
  {Benacquista}, {Cornish}, {Crowder}, {Larson}, {Plagnol}, {Porter},
  {Vallisneri}, {Vecchio} et~al.}}]{2008CQGra..25r4026B}
\bibinfo{author}{\bibfnamefont{S.}~\bibnamefont{{Babak}}},
  \bibinfo{author}{\bibfnamefont{J.~G.} \bibnamefont{{Baker}}},
  \bibinfo{author}{\bibfnamefont{M.~J.} \bibnamefont{{Benacquista}}},
  \bibinfo{author}{\bibfnamefont{N.~J.} \bibnamefont{{Cornish}}},
  \bibinfo{author}{\bibfnamefont{J.}~\bibnamefont{{Crowder}}},
  \bibinfo{author}{\bibfnamefont{S.~L.} \bibnamefont{{Larson}}},
  \bibinfo{author}{\bibfnamefont{E.}~\bibnamefont{{Plagnol}}},
  \bibinfo{author}{\bibfnamefont{E.~K.} \bibnamefont{{Porter}}},
  \bibinfo{author}{\bibfnamefont{M.}~\bibnamefont{{Vallisneri}}},
  \bibinfo{author}{\bibfnamefont{A.}~\bibnamefont{{Vecchio}}},
  \bibnamefont{et~al.}, \bibinfo{journal}{Classical and Quantum Gravity}
  \textbf{\bibinfo{volume}{25}}, \bibinfo{eid}{184026} (\bibinfo{year}{2008}),
  \eprint{0806.2110}.

\bibitem[{\citenamefont{{Gair} et~al.}(2011)\citenamefont{{Gair}, {Mandel},
  {Miller}, and {Volonteri}}}]{2011GReGr..43..485G}
\bibinfo{author}{\bibfnamefont{J.~R.} \bibnamefont{{Gair}}},
  \bibinfo{author}{\bibfnamefont{I.}~\bibnamefont{{Mandel}}},
  \bibinfo{author}{\bibfnamefont{M.~C.} \bibnamefont{{Miller}}},
  \bibnamefont{and}
  \bibinfo{author}{\bibfnamefont{M.}~\bibnamefont{{Volonteri}}},
  \bibinfo{journal}{General Relativity and Gravitation}
  \textbf{\bibinfo{volume}{43}}, \bibinfo{pages}{485} (\bibinfo{year}{2011}),
  \eprint{0907.5450}.

\bibitem[{\citenamefont{{Fotopoulos} and {LIGO Scientific
  Collaboration}}(2008)}]{2008JPhCS.122a2032F}
\bibinfo{author}{\bibfnamefont{N.~V.} \bibnamefont{{Fotopoulos}}}
  \bibnamefont{and} \bibinfo{author}{\bibnamefont{{LIGO Scientific
  Collaboration}}}, \bibinfo{journal}{Journal of Physics Conference Series}
  \textbf{\bibinfo{volume}{122}}, \bibinfo{eid}{012032} (\bibinfo{year}{2008}),
  \eprint{0801.3429}.

\bibitem[{\citenamefont{{Belczynski} et~al.}(2002)\citenamefont{{Belczynski},
  {Kalogera}, and {Bulik}}}]{2002ApJ...572..407B}
\bibinfo{author}{\bibfnamefont{K.}~\bibnamefont{{Belczynski}}},
  \bibinfo{author}{\bibfnamefont{V.}~\bibnamefont{{Kalogera}}},
  \bibnamefont{and} \bibinfo{author}{\bibfnamefont{T.}~\bibnamefont{{Bulik}}},
  \bibinfo{journal}{\apj} \textbf{\bibinfo{volume}{572}}, \bibinfo{pages}{407}
  (\bibinfo{year}{2002}), \eprint{astro-ph/0111452}.

\bibitem[{\citenamefont{{Belczynski} et~al.}(2008)\citenamefont{{Belczynski},
  {Kalogera}, {Rasio}, {Taam}, {Zezas}, {Bulik}, {Maccarone}, and
  {Ivanova}}}]{2008ApJS..174..223B}
\bibinfo{author}{\bibfnamefont{K.}~\bibnamefont{{Belczynski}}},
  \bibinfo{author}{\bibfnamefont{V.}~\bibnamefont{{Kalogera}}},
  \bibinfo{author}{\bibfnamefont{F.~A.} \bibnamefont{{Rasio}}},
  \bibinfo{author}{\bibfnamefont{R.~E.} \bibnamefont{{Taam}}},
  \bibinfo{author}{\bibfnamefont{A.}~\bibnamefont{{Zezas}}},
  \bibinfo{author}{\bibfnamefont{T.}~\bibnamefont{{Bulik}}},
  \bibinfo{author}{\bibfnamefont{T.~J.} \bibnamefont{{Maccarone}}},
  \bibnamefont{and}
  \bibinfo{author}{\bibfnamefont{N.}~\bibnamefont{{Ivanova}}},
  \bibinfo{journal}{Astrophys. J. Sup.} \textbf{\bibinfo{volume}{174}},
  \bibinfo{pages}{223} (\bibinfo{year}{2008}), \eprint{astro-ph/0511811}.

\bibitem[{\citenamefont{{Belczynski} et~al.}(2010)\citenamefont{{Belczynski},
  {Dominik}, {Bulik}, {O'Shaughnessy}, {Fryer}, and
  {Holz}}}]{2010ApJ...715L.138B}
\bibinfo{author}{\bibfnamefont{K.}~\bibnamefont{{Belczynski}}},
  \bibinfo{author}{\bibfnamefont{M.}~\bibnamefont{{Dominik}}},
  \bibinfo{author}{\bibfnamefont{T.}~\bibnamefont{{Bulik}}},
  \bibinfo{author}{\bibfnamefont{R.}~\bibnamefont{{O'Shaughnessy}}},
  \bibinfo{author}{\bibfnamefont{C.}~\bibnamefont{{Fryer}}}, \bibnamefont{and}
  \bibinfo{author}{\bibfnamefont{D.~E.} \bibnamefont{{Holz}}},
  \bibinfo{journal}{\apj} \textbf{\bibinfo{volume}{715}}, \bibinfo{pages}{L138}
  (\bibinfo{year}{2010}), \eprint{1004.0386}.

\bibitem[{\citenamefont{{Dominik} et~al.}(2012)\citenamefont{{Dominik},
  {Belczynski}, {Fryer}, {Holz}, {Berti}, {Bulik}, {Mandel}, and
  {O'Shaughnessy}}}]{2012ApJ...759...52D}
\bibinfo{author}{\bibfnamefont{M.}~\bibnamefont{{Dominik}}},
  \bibinfo{author}{\bibfnamefont{K.}~\bibnamefont{{Belczynski}}},
  \bibinfo{author}{\bibfnamefont{C.}~\bibnamefont{{Fryer}}},
  \bibinfo{author}{\bibfnamefont{D.~E.} \bibnamefont{{Holz}}},
  \bibinfo{author}{\bibfnamefont{E.}~\bibnamefont{{Berti}}},
  \bibinfo{author}{\bibfnamefont{T.}~\bibnamefont{{Bulik}}},
  \bibinfo{author}{\bibfnamefont{I.}~\bibnamefont{{Mandel}}}, \bibnamefont{and}
  \bibinfo{author}{\bibfnamefont{R.}~\bibnamefont{{O'Shaughnessy}}},
  \bibinfo{journal}{\apj} \textbf{\bibinfo{volume}{759}}, \bibinfo{eid}{52}
  (\bibinfo{year}{2012}), \eprint{1202.4901}.

\bibitem[{\citenamefont{{Kowalska} et~al.}(2012)\citenamefont{{Kowalska},
  {Regimbau}, {Bulik}, {Dominik}, and {Belczynski}}}]{2012arXiv1205.4621K}
\bibinfo{author}{\bibfnamefont{I.}~\bibnamefont{{Kowalska}}},
  \bibinfo{author}{\bibfnamefont{T.}~\bibnamefont{{Regimbau}}},
  \bibinfo{author}{\bibfnamefont{T.}~\bibnamefont{{Bulik}}},
  \bibinfo{author}{\bibfnamefont{M.}~\bibnamefont{{Dominik}}},
  \bibnamefont{and}
  \bibinfo{author}{\bibfnamefont{K.}~\bibnamefont{{Belczynski}}},
  \bibinfo{journal}{ArXiv e-prints}  (\bibinfo{year}{2012}),
  \eprint{1205.4621}.

\bibitem[{\citenamefont{{Hopkins} and {Beacom}}(2006)}]{2006ApJ...651..142H}
\bibinfo{author}{\bibfnamefont{A.~M.} \bibnamefont{{Hopkins}}}
  \bibnamefont{and} \bibinfo{author}{\bibfnamefont{J.~F.}
  \bibnamefont{{Beacom}}}, \bibinfo{journal}{\apj}
  \textbf{\bibinfo{volume}{651}}, \bibinfo{pages}{142} (\bibinfo{year}{2006}),
  \eprint{astro-ph/0601463}.

\bibitem[{\citenamefont{{Buonanno} et~al.}(2009)\citenamefont{{Buonanno},
  {Iyer}, {Ochsner}, {Pan}, and {Sathyaprakash}}}]{2009PhRvD..80h4043B}
\bibinfo{author}{\bibfnamefont{A.}~\bibnamefont{{Buonanno}}},
  \bibinfo{author}{\bibfnamefont{B.~R.} \bibnamefont{{Iyer}}},
  \bibinfo{author}{\bibfnamefont{E.}~\bibnamefont{{Ochsner}}},
  \bibinfo{author}{\bibfnamefont{Y.}~\bibnamefont{{Pan}}}, \bibnamefont{and}
  \bibinfo{author}{\bibfnamefont{B.~S.} \bibnamefont{{Sathyaprakash}}},
  \bibinfo{journal}{\prd} \textbf{\bibinfo{volume}{80}}, \bibinfo{eid}{084043}
  (\bibinfo{year}{2009}), \eprint{0907.0700}.

\bibitem[{\citenamefont{{Allen} and {Romano}}(1999)}]{1999PhRvD..59j2001A}
\bibinfo{author}{\bibfnamefont{B.}~\bibnamefont{{Allen}}} \bibnamefont{and}
  \bibinfo{author}{\bibfnamefont{J.~D.} \bibnamefont{{Romano}}},
  \bibinfo{journal}{\prd} \textbf{\bibinfo{volume}{59}}, \bibinfo{eid}{102001}
  (\bibinfo{year}{1999}), \eprint{gr-qc/9710117}.

\bibitem[{\citenamefont{{Flanagan}}(1993)}]{1993PhRvD..48.2389F}
\bibinfo{author}{\bibfnamefont{E.~E.} \bibnamefont{{Flanagan}}},
  \bibinfo{journal}{\prd} \textbf{\bibinfo{volume}{48}}, \bibinfo{pages}{2389}
  (\bibinfo{year}{1993}), \eprint{astro-ph/9305029}.

\bibitem[{\citenamefont{{Christensen}}(1992)}]{1992PhRvD..46.5250C}
\bibinfo{author}{\bibfnamefont{N.}~\bibnamefont{{Christensen}}},
  \bibinfo{journal}{\prd} \textbf{\bibinfo{volume}{46}}, \bibinfo{pages}{5250}
  (\bibinfo{year}{1992}).

\bibitem[{\citenamefont{{Meacher} et~al.}(2014)\citenamefont{{Meacher},
  {Thrane}, and {Regimbau}}}]{duncan}
\bibinfo{author}{\bibfnamefont{D.}~\bibnamefont{{Meacher}}},
  \bibinfo{author}{\bibfnamefont{E.}~\bibnamefont{{Thrane}}}, \bibnamefont{and}
  \bibinfo{author}{\bibfnamefont{T.}~\bibnamefont{{Regimbau}}}
  (\bibinfo{year}{2014}), \eprint{astro-ph/1402.6231}.

\end{thebibliography}
\end{document}